\documentstyle[prl,aps,floats,epsf]{revtex}
\begin{document}
\draft

\twocolumn[\hsize\textwidth\columnwidth\hsize\csname @twocolumnfalse\endcsname

\title{High Temperature Superconductivity and Charge Segregation in a 
Model with \\
Strong Long-Range Electron-Phonon and Coulomb Interactions}

\author{A.S. Alexandrov$^{1}$ and P.E. Kornilovitch $^{2}$}
\address{$^{1}$ Department of Physics, Loughborough University, 
Loughborough LE11 3TU, UK \\
$^{2}$ Hewlett Packard Labs, 1501 Page Mill Road, Palo Alto, 
CA 94304, USA}

\date{\today}

\maketitle

\begin{abstract}

An analytical method of studying strong long-range electron-phonon and Coulomb
interactions in complex lattices is presented.  The method is applied
to a perovskite layer with anisotropic coupling of holes to the vibrations
of apical atoms.  Depending on the relative strength of the polaronic shift, 
$E_p$, and the inter-site Coulomb repulsion, $V_c$, the system is either
a polaronic Fermi liquid, $V_c > 1.23 \, E_p$, a bipolaronic 
superconductor,  $1.16\, E_p < V_c < 1.23 \, E_p$, or a charge segregated
insulator, $V_c < 1.16\, E_p$.  In the superconducting window, the carriers
are mobile bipolarons with a remarkably low effective mass.  The model 
describes the key features of the underdoped superconducting cuprates. 
 
\end{abstract}


\vskip2pc]

\narrowtext

There is clear experimental \cite{guo,shen,tim,ega,mul2} and 
theoretical \cite{alemot,dev,allen,gor,alebra,bis2,feh,zey,bon2,aub2} 
evidence for strong electron-phonon (el-ph) interaction in high-$T_{c}$ 
cuprates.  Electron correlations are also important in shaping the 
Mott-Hubbard insulating state of parent undoped compounds \cite{and}. 
The theory of high-$T_c$ cuprates must treat both interactions on equal 
footing as was suggested some time ago \cite{alemot}.  In recent years 
many publications addressed the fundamental problem of competing 
el-ph and Coulomb interactions in the framework of the
Holstein-Hubbard model \cite{bis2,feh,zey,bon2,aub2} 
where both interactions are short-range (on-site).  The mass of 
bipolaronic carriers in this model is very large and the critical
temperature is suppressed down to a kelvin scale.  However, in the
cuprates the screening is poor so that the el-ph interaction necessarily
has to be long range.  Motivated by this fact, we have proposed that
a long range Fr\"ohlich, rather than short range Holstein, interaction
should be the adequate model for the cuprates \cite{ale2,alekor}.  
A small polaron with the Fr\"ohlich interaction was discussed long
time ago \cite{eagles}.  Analytical \cite{ale2} and exact
Monte-Carlo \cite{alekor} studies of the simple chain and plane models 
with a long-range el-ph coupling revealed a several order lower effective
mass of this polaron than that of the small Holstein polaron.  
Later the polaron and 
bipolaron cases of the chain model were analyzed in more detail in 
Refs.\cite{feh2} and \cite{bon}  confirming low masses of both types
of carriers.  Qualitatively, a long-range el-ph interaction results in a 
lighter mass because the extended lattice deformation changes gradually 
as the carrier tunnels through the lattice. 

In this Letter we study a realistic multi-polaron model of the 
copper-oxygen perovskite layer, a major structural unit of the 
HTSC compounds.  The model includes
the infinite on-site repulsion (Hubbard $U$ term), long-range inter-hole
Coulomb repulsion $V_c$, and long-range Fr\"ohlich interaction between
in-plane holes and apical oxygens.  We find that within a certain window 
of $V_c$ the holes form intersite bipolarons with a remarkably low mass.
The bipolarons repel and the whole system is a superconductor with a high
critical temperature.  At large $V_c$, the system is a  
polaronic Fermi-liquid, and at small $V_c$ it is a charge segregated 
insulator.
   
To deal with the considerable complexity of the model we first describe a 
theoretical approach that makes the analysis of complex lattices simple
in the strong coupling limit.  The model Hamiltonian explicitly includes 
long-range electron-phonon and Coulomb interactions as well as 
kinetic and deformation energies.  An implicitly present infinite Hubbard 
term prohibits double occupancy and removes the need to distinguish  
fermionic spins.  Introducing spinless fermion, $c_{\bf n}$, and 
phonon, $d_{{\bf m}\alpha }$, operators the Hamiltonian is written as 
\begin{eqnarray}
H = & - & \sum_{\bf n \neq n'} 
\left[ T({\bf n-n'}) c_{\bf n}^{\dagger } c_{\bf n'}
- V_{c}({\bf n-n'}) c_{\bf n}^{\dagger} c_{\bf n}
c_{\bf n'}^{\dagger } c_{\bf n'} \right]  \nonumber \\ 
& - & \omega  \sum_{\bf n m}  g_{\alpha}({\bf m-n}) 
({\bf e}_{{\bf m}\alpha } \cdot {\bf u}_{\bf m-n})
c_{\bf n}^{\dagger } c_{\bf n}
(d_{{\bf m}\alpha}^{\dagger}+d_{{\bf m}\alpha })  \nonumber \\
& + & \omega \sum_{{\bf m}\alpha} \left( d_{{\bf m}\alpha }^{\dagger} 
d_{{\bf m}\alpha }+\frac{1}{2} \right) .
\label{one}
\end{eqnarray}
Here ${\bf e}_{{\bf m} \alpha}$ is the polarization vector of $\alpha$th
vibration coordinate at site ${\bf m}$, 
${\bf u}_{\bf m-n} \equiv ({\bf m-n})/|{\bf m-n}|$ is the unit vector in 
the direction from electron ${\bf n}$ to ion ${\bf m}$, 
and $g_{\alpha}({\bf m-n)}$ is a dimensionless el-ph coupling function. 
[$g_{\alpha}({\bf m-n)}$ is proportional to the {\em force} acting between
${\bf m}$ and ${\bf n}$.]   We assume that all the phonon modes are 
dispersionless with frequency $\omega$ and that the electrons do not 
interact with displacements of their own atoms, $g_{\alpha}(0) \equiv 0$.  
We also use $\hbar = 1$ throughout the paper. 

In the limit of strong el-ph interaction it is convenient to
perform the Lang-Firsov canonical transformation \cite{lan} . 
Introducing $S = \sum_{{\bf mn}\alpha } g_{\alpha}({\bf m-n})
({\bf e}_{{\bf m}\alpha} \cdot {\bf u}_{\bf m-n}) 
c_{\bf n}^{\dagger } c_{\bf n}
( d_{{\bf m}\alpha }^{\dagger } - d_{{\bf m}\alpha })$ 
one obtains a transformed Hamiltonian without an explicit el-ph term 
\begin{eqnarray}
\tilde{H} & = & e^{-S} H e^{S} = 
-\sum_{{\bf n \neq n'}} \hat{\sigma}_{\bf nn'} c_{\bf n}^{\dagger }c_{\bf n'}
+ \omega \sum_{{\bf m} \alpha} \left( d_{{\bf m}\alpha }^{\dagger } 
d_{{\bf m}\alpha}+\frac{1}{2} \right) \nonumber \\
& + & \sum_{\bf n \neq n'} v({\bf n-n'}) c_{\bf n}^{\dagger } c_{\bf n}
c_{\bf n'}^{\dagger } c_{\bf n'}  
- E_{p} \sum_{\bf n} c_{\bf n}^{\dagger } c_{\bf n} .
\label{two}
\end{eqnarray}
The last term describes the energy which polarons gain due to 
el-ph interaction.  $E_{p}$ is the familiar polaronic (Franc-Condon) 
level shift 
\begin{equation}
E_{p} = \omega \sum_{{\bf m} \alpha} g_{\alpha}^{2}({\bf m-n})
({\bf e}_{{\bf m}\alpha }\cdot {\bf u}_{\bf m-n})^{2} ,
\label{three}
\end{equation}
which we assume to be independent of ${\bf n}$.  
$E_p$ is a natural measure of the strength of the el-ph interaction. 
The third term in Eq.(\ref{two}) is the polaron-polaron interaction: 
\begin{equation}
v({\bf n-n'}) = V_{c}({\bf n-n'}) - V_{\rm pa}({\bf n-n'}) , 
\label{five}
\end{equation}
\begin{eqnarray}
V_{\rm pa}({\bf n-n'}) = 
2\omega \sum_{{\bf m} \alpha}
g_{\alpha}({\bf m-n}) g_{\alpha}({\bf m-n'}) \times \nonumber \\ 
({\bf e}_{{\bf m}\alpha }\cdot {\bf u}_{\bf m-n})
({\bf e}_{{\bf m}\alpha }\cdot {\bf u}_{\bf m-n'}),
\label{fiveone}
\end{eqnarray}
where $V_{\rm pa}$ is the inter-polaron {\em attraction} due to joint 
interaction with the same vibrating atoms.  Finally, the first term in 
Eq.(\ref{two}) contains a transformed hopping operator 
$\hat{\sigma}_{\bf nn'}$:
\begin{eqnarray}
\hat{\sigma}_{\bf nn'} & = & T({\bf n-n'}) 
\exp \left[ \sum_{{\bf m} \alpha } 
\left[ g_{\alpha}({\bf m-n})({\bf e}_{{\bf m}\alpha } 
\cdot {\bf u}_{\bf m-n}) \right. \right. \nonumber \\
 & - & \left. \left.
g_{\alpha}({\bf m-n'})({\bf e}_{{\bf m}\alpha } 
\cdot {\bf u}_{\bf m-n'})\right]
(d_{{\bf m}\alpha }^{\dagger } - d_{{\bf m}\alpha}) \right] .
\label{four}
\end{eqnarray}
At large $E_p/T({\bf n-n'})$ this term is a perturbation.  
In the first order of the strong coupling perturbation theory 
\cite{alemot}, $\hat{\sigma}_{\bf nn'}$ 
should be averaged over phonons because there is no coupling between 
polarons and phonons in the unperturbed Hamiltonian [the last three terms 
in Eq.(\ref{two})].  For temperatures lower than $\omega$, the result is 
\begin{equation}
t({\bf n-n'}) \equiv \left\langle \hat{\sigma}_{{\bf nn'}}\right\rangle_{ph}
= T({\bf n-n'}) \exp [-G^{2}({\bf n-n'})],  
\label{six}
\end{equation}
\begin{eqnarray}
G^{2}({\bf n-n'}) = \sum_{{\bf m} \alpha} g_{\alpha} ({\bf m-n}) 
({\bf e}_{{\bf m}\alpha } \cdot {\bf u}_{\bf m-n}) \times
\nonumber \\
\left[ g_{\alpha}({\bf m-n}) ({\bf e}_{{\bf m}\alpha } 
\cdot {\bf u}_{\bf m-n})
- g_{\alpha} ({\bf m-n'}) ({\bf e}_{{\bf m}\alpha } 
\cdot {\bf u}_{\bf m-n'}) \right] .
\label{seven}
\end{eqnarray}
By comparing Eqs.(\ref{three}), (\ref{fiveone}), and (\ref{seven}), 
the mass renormalization exponents can be expressed via $E_p$ and
$V_{\rm pa}$ as follows
\begin{equation}
G^2({\bf n-n'}) = \frac{1}{\omega} 
\left( E_p - \frac{1}{2}V_{\rm pa}({\bf n-n'}) \right) .
\label{sevenone}
\end{equation}
This is the simplest way to calculate $G^2$ and (bi)polaron masses
once the `static' parameters  $E_p$ and $V_{\rm pa}$ are known.

It is easy to see from the above equations that the long-range el-ph
interaction increases $E_p$ and $V_{\rm pa}$ but {\em reduces} $G^2$ 
(when measured in natural units of $E_p/\omega$).
Thus polarons get tighter and at the same time lighter.  Bipolarons
form when $V_{\rm pa}$ exceeds $V_c$ and they are relatively light too.  
We note that the Holstein model is the limiting case with the highest
possible $G^2 = E_p/\omega$.  In this respect, the Holstein model
is {\em not} a typical el-ph model. 

\begin{figure}[t]
\vspace{-1.cm}
\begin{center}
\leavevmode
\hbox{
\epsfxsize=8.4cm
\epsffile{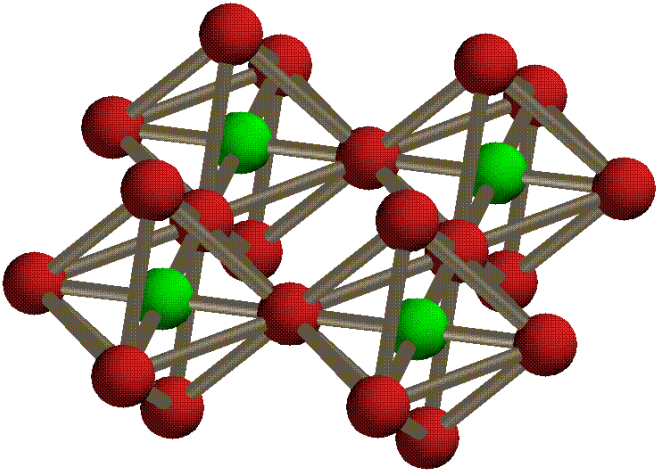}
}
\end{center}
\vspace{-4.5cm}
\caption{
Four octahedra of the copper-oxygen perovskite layer.  Holes reside on
the in-plane oxygens but interact with apical oxygens.
}
\label{fig1}
\end{figure}

To obtain an analytical description of the multi-polaron 
system we restrict our consideration to the strong coupling case 
$t \leq |v|$.  In this regime the polaron kinetic energy is
the smallest energy and thus can be treated as a perturbation.  The
system is adequately described by a purely polaronic model:
\begin{equation}
H_{p} = H_{0} + H_{\rm pert},
\label{eight}
\end{equation}
\begin{equation}
H_{0}=-E_{p}\sum_{{\bf n}}c_{{\bf n}}^{\dagger }c_{{\bf n}}+
   \sum_{{\bf  n \neq n'}} v({\bf n-n'})c_{{\bf n}}^{\dagger }
c_{{\bf n}} c_{{\bf n'}}^{\dagger} c_{\bf n'}, 
\label{nine}
\end{equation}
\begin{equation}
H_{\rm pert} = - \sum_{{\bf n \neq n'}} t({\bf n-n'})
c_{\bf n}^{\dagger } c_{\bf n'}.
\label{ten} 
\end{equation}
The many-particle ground state of $H_{0}$ depends on the sign of the
polaron-polaron interaction, the carrier density, and the lattice geometry.
Here we consider a two dimensional lattice of ideal octahedra that
can be regarded as a simplified model of the copper-oxygen perovskite 
layer, see Figure~\ref{fig1}.  The lattice period is $a=1$ and the distance 
between the apical sites and the central plane is $h=a/2 = 0.5$. 
All in-plane atoms, both copper and oxygen, are static but apical
oxygens are independent three-dimensional isotropic harmonic oscillators.
Because of poor screening the hole-apical interaction is purely
Coulombic, $g_{\alpha}({\bf m-n}) = \kappa_{\alpha}/|{\bf m-n}|^2$,
$\alpha = x,y,z$.  To account for the experimental fact that $z$-polarized
phonons couple to the holes stronger than the others \cite{tim}    
we choose $\kappa_x = \kappa_y = \kappa_z/\sqrt{2}$.  The direct
hole-hole repulsion is $V_c({\bf n-n'}) = \frac{V_c/\sqrt{2}}{|{\bf n-n'}|}$
so that the repulsion between two holes in the NN configuration is $V_c$. 
We also include the bare nearest neighbor (NN) hopping $T_{NN}$, 
the next nearest neighbor (NNN) hopping across copper $T_{NNN}$ 
and the NNN hopping between octahedra $T'_{NNN}$.  

According to Eq.(\ref{three}), the polaron shift is given by the
lattice sum (after summation over polarizations):
\begin{eqnarray}
E_p & = & 2 \kappa^2_x \omega \sum_{\bf m} 
\left( \frac{1}{|{\bf m-n}|^{4}} + \frac{h^2}{|{\bf m-n}|^{6}} \right) 
\nonumber \\
    & = & 31.15 \, \kappa^2_x \omega ,
\label{eleven}
\end{eqnarray}
where the factor 2 accounts for the two layers of apical sites.
[For reference, Cartesian coordinates are 
${\bf n} = (n_x+1/2, n_y+1/2, 0)$, ${\bf m} = (m_x, m_y, h)$;
$n_x,n_y,m_x,m_y$ being integers.] 
The polaron-polaron attraction is 
\begin{equation}
V_{\rm pa}({\bf n-n'}) = 4 \omega \kappa^2_x \sum_{\bf m}
\frac{h^2 + ({\bf m-n'}) \cdot({\bf m-n}) } 
{|{\bf m-n'}|^3 |{\bf m-n}|^3} .
\label{twelve}
\end{equation}
Performing lattice summations for the NN, NNN, and NNN$'$ configurations
one finds $V_{\rm pa} = 1.23\, E_p, 0.80 \, E_p$, and $0.82 \, E_p$,
respectively.  Substituting these results in Eqs.(\ref{five}) and 
(\ref{sevenone}) we obtain the full inter-polaron interaction:
$v_{NN} = V_c - 1.23\, E_p$, $v_{NNN} = \frac{V_c}{\sqrt{2}} - 0.80\, E_p$,
$v'_{NNN} = \frac{V_c}{\sqrt{2}} - 0.82\, E_p$,   
and the mass renormalization exponents:
$G^2_{NN} = 0.38 (E_p/\omega)$, $G^2_{NNN} = 0.60 (E_p/\omega)$ and 
$G'^2_{NNN} = 0.59 (E_p/\omega)$.

Let us now discuss different regimes of the model.  At $V_c > 1.23 \, E_p$,
no bipolarons are formed and the systems is a polaronic Fermi liquid.
The polarons tunnel in the {\em square} lattice with NN hopping
$t = T_{NN} \exp ( -0.38 E_p/\omega)$ and NNN hopping 
$t' = T_{NNN} \exp ( -0.60 E_p/\omega)$.  [Since 
$G^2_{NNN} \approx G'^2_{NNN}$ one can neglect the difference between
NNN hoppings within and between the octahedra.]  The single polaron
spectrum is therefore    
\begin{eqnarray}
E_1({\bf k})= - E_p & -   & 2t'[\cos k_x +\cos k_y] \nonumber \\ 
                    & \pm & 4t\cos(k_x/2)\cos(k_y/2) .
\label{fifteen}
\end{eqnarray}
The polaron mass is $m^{\ast} = 1/(t+2t')$.  Since in general $t > t'$,
the mass is mostly determined by the NN hopping amplitude $t$. 

\begin{figure}[t]
\vspace{-1.cm}
\begin{center}
\leavevmode
\hbox{
\epsfxsize=8.4cm
\epsffile{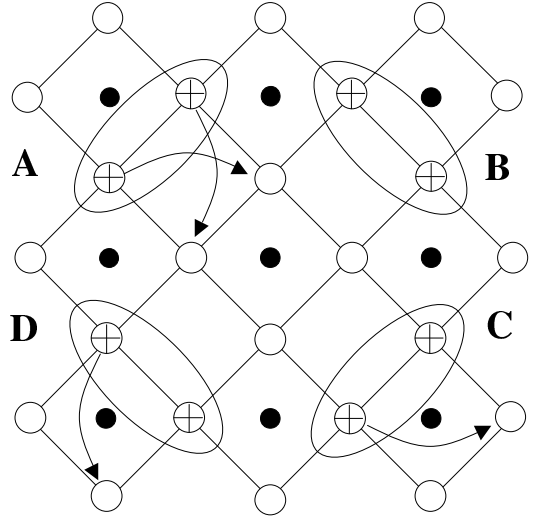}
}
\end{center}
\caption{
Top view on the perovskite layer.  The apical sites are not shown.
The four bipolaron configurations $A$, $B$, $C$, and $D$ all have
the same energy.  Some possible single-polaron hoppings $t'$ are 
indicated by arrows.  Note that the bipolaron movement is first
order in $t'$.  
}
\label{fig2}
\end{figure}

If $V_c < 1.23 \,E_p$, then intersite NN bipolarons form.  The bipolarons 
tunnel in the plane via four resonating (degenerate) configurations
$A$, $B$, $C$, and $D$, see Figure~\ref{fig2}.  In the first order in 
$H_{\rm pert}$ one should retain only these lowest energy configurations 
and discard all the processes that involve configurations with higher 
energies.  The result of such a projection is the bipolaronic Hamiltonian
\begin{eqnarray}
H_{b}&=&(V_{c}-3.23 \, E_{p}) \sum_{\bf l}
                    [A^{\dagger}_{\bf l} A_{\bf l}
                   + B^{\dagger}_{\bf l} B_{\bf l}
                   + C^{\dagger}_{\bf l} C_{\bf l}
                   + D^{\dagger}_{\bf l} D_{\bf l}]
\nonumber \\
& - & t'\sum_{\bf l}[A^{\dagger}_{\bf l} B_{\bf l}
                   + B^{\dagger}_{\bf l} C_{\bf l}
                   + C^{\dagger}_{\bf l} D_{\bf l}
                   + D^{\dagger}_{\bf l} A_{\bf l} + {\rm h.c.}] \nonumber \\
& - &t'\sum_{\bf n} [A^{\dagger}_{\bf l-x} B_{\bf l}
                   + B^{\dagger}_{\bf l+y} C_{\bf l} \nonumber \\
&   &\makebox[0.5cm]{}
                   + C^{\dagger}_{\bf l+x} D_{\bf l}
                   + D^{\dagger}_{\bf l-y} A_{\bf l} 
+ {\rm h.c.} ] ,
\label{sixteen}
\end{eqnarray}
where ${\bf l}$ numbers octahedra rather than individual sites,
${\bf x} = (1,0)$, and ${\bf y} = (0,1)$.
A Fourier transformation and diagonalization of a $4 \times 4$ matrix
yields the bipolaron spectrum:
\begin{equation}
E_{2}({\bf k}) = V_{c} - 3.23 E_p \pm 2t'[\cos (k_x/2)\pm \cos (k_y/2)].
\label{seventeen}
\end{equation}
There are four bipolaronic subbands combined in a band of width $8t'$. 
The effective mass of the lowest band is $m^{\ast\ast} = 2/t'$.  
The bipolaron binding energy is 
$\Delta = 2E_1(0) - E_2(0) = 1.23 E_p - V_c - 8t - 4t'$.  

We have to emphasize that the bipolaron moves already in the {\em first} 
order in polaron hopping.  This remarkable property is entirely due to the 
strong on-site repulsion and long-range electron-phonon interaction 
that leads to a non-trivial connectivity of the lattice.  This situation 
is unlike all other models studied previously.  [Usually the bipolaron
moves only in the second order in polaron hopping and therefore is
very heavy.]  In our model, this fact combines with a weak renormalization
of $t'$ yielding a {\em superlight} bipolaron with mass 
$m^{\ast\ast} \propto \exp(0.60\, E_p/\omega)$.  We recall that
in the Holstein model $m^{\ast\ast} \propto \exp(2 E_p/\omega)$. 
Thus the mass of the Fr\"ohlich bipolaron scales approximately
as the {\em cubic root} of that of the Holstein one.

At even stronger el-ph interaction, $V_c < 1.16 E_p$, NNN bipolarons
become stable.  More importantly, holes can now form 3- and 4-particle
clusters.  Such clusters do not have resonant states and remain immobile
in the first order in polaron hopping.  The system quickly becomes
a charge segregated insulator.     

The superconductivity window that we have found, 
$1.16 E_p < V_c < 1.23 E_p$, is quite narrow.  This indicates that 
the superconducting state in such systems is a subtle phenomenon
which requires a fine balance between electronic and ionic interactions.  
Too strong el-ph interaction leads to clustering, while
too weak interaction cannot bind the carriers and the superconductivity
is at best of BCS type.  These considerations may provide
additional insight into the uniqueness of one particular structure, 
the copper-oxygen perovskite layer, to HTSC.  It also follows from
our model that superconductivity should be very sensitive to any 
external factor that affects the balance between $V_c$ and $E_p$.
For instance, pressure changes the octahedra geometry and hence
$E_p$ and $V_{\rm pa}$.  Chemical doping enhances internal screening
and consequently reduces $E_p$.      

We now assume that the superconductivity condition is satisfied and
show that our `Fr\"ohlich-Coulomb' model possesses many key properties 
of the underdoped cuprates.  The bipolaron binding energy $\Delta$ 
should manifest itself as a normal state pseudogap with size of 
approximately half of $\Delta$ \cite{alemot}.  Such a pseudogap is 
indeed observed in many cuprates.  There should be a strong isotope 
effect on the (bi)polaron mass because 
$t,t' \propto \exp(-{\rm const} \sqrt{M})$.
Therefore the replacement of O$^{16}$ by O$^{18}$ increases the carrier 
mass \cite{ale3}.  Such an effect has been observed in the
London penetration depth of the isotope-substituted samples
\cite{guo}.  The mass isotope exponent, $\alpha_m=d\ln
m^{\ast\ast}/d\ln M$, was found to be as large as $\alpha_m = 0.8$ in
La$_{1.895}$Sr$_{0.105}$CuO$_4$.  Our theoretical exponent is
$\alpha_m=0.3 E_{p}/\omega$, so that the bipolaron mass enhancement
factor is $\exp(0.6E_p/\omega) \simeq 5$ in this material.  With the bare 
hopping integral $T_{NNN}=0.2$ eV we obtain the in-plane bipolaron mass 
$m^{\ast \ast} \simeq 10 m_e$.  Calculated with this value the in-plane 
London penetration depth, 
$\lambda_{ab}=[m^{\ast \ast}/8\pi ne^{2}]^{1/2}\simeq 316$ nm 
($n$ the hole density) agrees well with the measured one 
$\lambda_{ab} \simeq 320$ nm.  Taking into account the c-axis tunneling 
of bipolarons, the critical temperature of their Bose-Einstein condensation 
can be expressed in terms of the experimentally measured in-plane and 
c-axis penetration depths, and the in-plane Hall constant $R_{H}$ as 
$T_c = 1.64 f \cdot (eR_H/\lambda_{ab}^4\lambda_{c}^2)^{1/3}$.  Here 
$f \approx 1$ and 
$T_c$, $eR_{H}$, and $\lambda$ are measured in K, cm$^3$
and cm, respectively \cite{alekab}.  Using the experimental
$\lambda_{ab} = 320$ nm, $\lambda_{c} = 4160$ nm, and 
$R_H = 4 \times 10^{-3}$ cm$^3$/C (just above $T_c$) one obtains 
$T_c = 31$ K in striking agreement with the experimental value $T_c=30$ K. 
The recent observation of the normal state diamagnetism in 
La$_{2-x}$Sr$_{x}$CuO$_{4}$ \cite{nat} also confirms 
the prediction of the bipolaron theory \cite{alekabden}.  
Many other features of the bipolaronic (super)conductor, e.g.,
the unusual upper critical field, electronic specific heat, optical and
tunneling spectra  match those of the cuprates (for a recent review, 
see Ref. \cite{aleedw}).  

In conclusion, we have studied a model with strong long-range 
electron-phonon and Coulomb interactions.  The model shows a 
reach phase diagram depending on the ratio of the inter-site Coulomb 
repulsion and the polaronic (Franc-Condon) level shift.  
The ground state is a polaronic Fermi (or Luttinger) liquid at large
Coulomb repulsions, a bipolaronic high-temperature superconductor at 
intermediate Coulomb repulsions, and a charge-segregated insulator 
at weak repulsion.  In the superconducting phase, inter-site bipolarons 
are remarkably light leading to a high critical temperature.
The model describes many properties of the superconducting cuprates. 

\bigskip

This work has been supported by EPSRC UK, grant R46977 (ASA), and
by DARPA (PEK).

\end{document}